\documentclass[sigconf]{acmart}

\usepackage{array}
\usepackage{listings}
\usepackage{url}
\usepackage{amsmath,amsfonts}
\usepackage{graphicx}
\usepackage{textcomp}
\usepackage{xcolor}
\usepackage{wrapfig}
\usepackage{multirow}
\usepackage{booktabs}
\usepackage{multirow}
\usepackage{float}
 \usepackage{tikz}

\usepackage{framed}
\usepackage[skip=3pt]{caption}
\usepackage{subcaption}
\usepackage{graphicx,xcolor} 
\usepackage[linesnumbered,ruled,vlined]{algorithm2e}
\usepackage{tikz}
\usepackage{nicefrac}

\usepackage{wrapfig}
\newcommand*\circled[1]{\tikz[baseline=(char.base)]{
            \node[shape=circle,fill,inner sep=0.5pt] (char) {\textcolor{white}{#1}};}}

\newcommand\blfootnote[1]{%
  \begingroup
  \renewcommand\thefootnote{}\footnote{#1}%
  \addtocounter{footnote}{-1}%
  \endgroup
}

\usepackage{pifont}
\newcommand{\cmark}{\ding{51}}%
\newcommand{\xmark}{\ding{55}}%

\usepackage{xcolor,colortbl}
\def\BibTeX{{\rm B\kern-.05em{\sc i\kern-.025em b}\kern-.08em
    T\kern-.1667em\lower.7ex\hbox{E}\kern-.125emX}}

\usepackage{tikz}

\AtBeginDocument{%
  \providecommand\BibTeX{{%
    \normalfont B\kern-0.5em{\scshape i\kern-0.25em b}\kern-0.8em\TeX}}}

\copyrightyear{2025}
\acmYear{2025}
\setcopyright{usgovmixed}
\acmConference[ICPP '25]{54th International Conference on Parallel Processing}{September 08--11, 2025}{San Diego, CA, USA}
\acmBooktitle{54th International Conference on Parallel Processing (ICPP '25), September 08--11, 2025, San Diego, CA, USA}
\acmPrice{}
\acmDOI{10.1145/3754598.3754651}
\acmISBN{979-8-4007-2074-1/25/09}

\makeatletter
\gdef\@copyrightpermission{
 \begin{minipage}{0.3\columnwidth}
 \href{https://creativecommons.org/licenses/by/4.0/}{\includegraphics[width=0.90\textwidth]{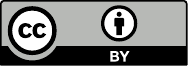}}
 \end{minipage}\hfill
 \begin{minipage}{0.7\columnwidth}
 \href{https://creativecommons.org/licenses/by/4.0/}{This work is licensed under a Creative Commons Attribution
International 4.0 License.}
 \end{minipage}
 \vspace{5pt}
}
\makeatother

\ccsdesc[500]{Computing methodologies~Concurrent algorithms}
\ccsdesc[500]{Computing methodologies~Massively parallel algorithms}
\ccsdesc[500]{Computing methodologies~Vector / streaming algorithms}

\begin{document}

\title[AMPED: Accelerating MTTKRP for Billion-Scale Sparse Tensor Decomposition on Multiple GPUs]{AMPED: \underline{A}ccelerating \underline{M}TTKRP for Billion-Scale S\underline{p}arse \\ T\underline{e}nsor \underline{D}ecomposition on Multiple GPUs}

\author{Sasindu Wijeratne}
\affiliation{%
  \institution{University of Southern California}
  \city{Los Angeles}
  \state{California}
  \country{USA}
}
\email{kangaram@usc.edu}

\author{Rajgopal Kannan}
\affiliation{%
  \institution{DEVCOM Army Research Office}
  \city{Los Angeles}
  \state{California}
  \country{USA}
}
\email{rajgopal.kannan.civ@army.mil}

\author{Viktor Prasanna}
\affiliation{%
  \institution{University of Southern California}
  \city{Los Angeles}
  \state{California}
  \country{USA}
}
\email{prasanna@usc.edu}

\begin{abstract}
Matricized Tensor Times Khatri-Rao Product (MTTKRP) is the computational bottleneck in sparse tensor decomposition. As real-world sparse tensors grow to billions of nonzeros, they increasingly demand higher memory capacity and compute throughput from hardware accelerators. In this work, we present AMPED, a multi-GPU parallel algorithm designed to accelerate MTTKRP on billion-scale sparse tensors. AMPED scales beyond the limits of a single GPU, meeting both the memory and performance requirements of large-scale workloads. We introduce a partitioning strategy combined with a dynamic load balancing scheme to distribute computation and minimize GPU idle time. On real-world billion-scale tensors, AMPED achieves a 5.1$\times$ geometric mean speedup in total execution time over state-of-the-art GPU baselines using 4 GPUs on a single CPU node.
\end{abstract}



\keywords{MTTKRP, multi-GPU, Tensor Decomposition}

\maketitle

\section{Introduction}
\blfootnote{\textbf{Distribution Statement A:} Approved for public release. Distribution is unlimited.}
Tensors provide a natural way to represent data with multiple dimensions. Tensor decomposition transforms tensors with higher dimensionalities to a reduced latent space that can be leveraged to learn salient features of the underlying data distribution. Different application domains, including machine learning~\cite{8884203}, signal processing~\cite{cichocki2015tensor}, and network analysis~\cite{fernandes2021tensor} have employed tensor decomposition to achieve superior performance compared to conventional approaches. Canonical Polyadic Decomposition (CPD)~\cite{hong2020generalized} has become the widely used tensor decomposition method, where the Matricized Tensor Times Khatri-Rao Product (MTTKRP) is the computational bottleneck~\cite{kolda2009tensor}.

Since real-world tensors are generally sparse, tensor formats only keep nonzero tensor elements to reduce memory consumption~\cite{9622851,alto_paper,10.1145/3543622.3573179}. There is a need to develop optimized sparse tensor data layouts that support the highly irregular data access patterns of MTTKRP while accessing input tensors and factor matrices~\cite{9622851,alto_paper,10.1145/3543622.3573179}.

Real-world tensors often exhibit irregular shapes and distributions of nonzero tensor elements, which pose significant challenges when performing MTTKRP computations on multiple GPUs. These challenges arise from irregular memory access patterns, load imbalance among many GPU streaming multiprocessors, and the synchronization overhead among GPUs.

The size of real-world tensors is increasing rapidly with recent advances in big data applications~\cite{cichocki2014era}. Such applications use tensors with billions of nonzero tensor elements (i.e., billion-scale tensors). Therefore, parallel algorithms that scale beyond a single GPU are required to perform MTTKRP~\cite{10.1145/3626183.3659980, 5318} on such large tensors. However, distributing the sparse tensor across multiple GPUs can lead to load imbalance, latency in data migration, and intermediate value communication among devices, which can result in additional overhead to execution time.

Recent works have proposed accelerating MTTKRP on CPU~\cite{alto_paper, 9820702, wijeratne2023dynasor, laukemann2024accelerating}, GPU~\cite{10.1145/3295500.3356216, 8821030, wijeratne2024sparse, 10.1145/3524059.3532363, 10740878}, and ASIC~\cite{9065579}. In CPU and ASIC, the external memory is large enough to maintain billion-scale tensors. Meanwhile, general-purpose GPUs encounter additional scalability challenges on a single device due to limited GPU global memory.

In prior work, multiple GPUs were used to accelerate MTTKRP on millions-scale sparse tensors~\cite{9835348}. Such implementations do not scale well on billion-scale sparse tensors because of (1) the significant idle time incurred by some GPUs due to workload imbalance across GPUs, (2) increased communication overhead due to the latency in sharing intermediate results across GPUs, (3) the synchronization overhead across GPUs, and (4) scheduling overhead of dynamic load-balancing schemes during execution.

Various efforts have been made to balance the workload of MTTKRP across streaming multiprocessors (SMs) of a single GPU~\cite{8821030, wijeratne2024sparse, 10.1145/3524059.3532363, li2018parti}. FLYCOO-GPU~\cite{wijeratne2024sparse} introduces a partitioning scheme that eliminates task dependencies between GPU SMs for each mode and minimizes workload imbalance among GPU SMs. Our work extends the FLYCOO-GPU~\cite{wijeratne2024sparse} partitioning scheme to a multi-GPU environment, addressing a new set of challenges: (1) eliminating data dependencies between tensor partitions to minimize GPU-GPU communication and synchronization overhead, (2) balancing the workload across GPUs to minimize the idle time for each GPU, and (3) avoiding host CPU computations when collecting and distributing input tensor partitions and factor matrices, since CPU computing power is significantly lower than GPUs.


\begin{table*}[ht!] 
\vspace{-2mm}
\centering
\caption{Summary of Related Work}
\begin{tabular}{|c|c|c|c|c|c|}\hline
Work & Number of tensor & Multi-GPU & Load-balancing & Support for & Task independent \\
 & copies required & support & & billion-scale tensors & partitioning across GPUs  \\\hline\hline
 AMPED (ours) & No. of modes & \cmark & \cmark & \cmark & \cmark \\\hline
 MM-CSF~\cite{8821030} & No. of modes & \xmark & \cmark & \xmark & \xmark$^{*}$ \\\hline
FLYCOO-GPU~\cite{wijeratne2024sparse} & 2 & \xmark & \cmark & \xmark & \xmark$^{*}$ \\\hline
 BLCO~\cite{10.1145/3524059.3532363} & 1 & \xmark & \xmark & \cmark & \xmark$^{*}$ \\\hline
 HPSPTM~\cite{9835348} & No. of modes & \cmark & \xmark & \xmark & \xmark \\\hline
 ParTI-GPU~\cite{li2018parti} & 1 & \xmark & \cmark & \xmark & \xmark$^{*}$ \\\hline
\end{tabular}

    \vspace{1mm}
    \hspace{-5mm} $*$: Only support single GPU
\label{tab:relatedwork}
\end{table*}

The key contributions of this work are:
\begin{itemize}
\item We introduce a novel parallel algorithm to perform MTTKRP on sparse tensors using multiple GPUs. The proposed algorithm achieves geometric mean speedups of 1.9$\times$, 2.3$\times$, and 3.3$\times$ in execution time while using 2, 3, and 4 GPUs, compared to a single GPU implementation on billion-scale tensors.

\item We propose a tensor partitioning scheme to distribute the tensor elements across multiple GPUs. The proposed partitioning scheme achieves a geometric mean speedup of 8.2$\times$ in total execution time compared with equally distributing the nonzero tensor elements across the GPUs.



\item Our work achieves a geometric mean speedup of 5.1$\times$ in total execution time on billion-scale tensors compared with the state-of-the-art GPU baselines.
\end{itemize}

\section{Background and Related Work}\label{background}

\subsection{Introduction to Tensors}\label{background:intro}\label{background:decomp}

\begin{table}[H]
\vspace{-3mm}
\caption{Notations}
\begin{center}
\begin{tabular}{|c|c|}
\hline
     \textbf{Symbol} & \textbf{Details} \\
     \hline
          $\circ$ & vector outer product \\
     $\otimes$ & Kronecker product \\
     $\odot$ & Khatri-Rao product \\
$\mathbf{A}$ & matrix \\
$\mathbf{a}$ & vector \\
     $a$ & scalar \\
     $\mathcal{X}$ & sparse tensor \\
     $\mathcal{X}_{(d)}$ & mode-$d$ matricization of $\mathcal{X}$ \\
     \hline
\end{tabular}
\label{table:notation}
\end{center}
\vspace{-3mm}
\end{table}

A tensor is a generalization of an array to multiple dimensions. In the simplest high-dimensional case, a tensor is a three-dimensional array, which can be visualized as a data cube. For a thorough review of tensors, refer to~\cite{kolda2009tensor}. Table~\ref{table:notation} summarizes the tensor notations.


\subsubsection{Tensor mode} In Tensor Decomposition, the number of dimensions of an input tensor is commonly called the number of tensor modes. For example, a vector can be seen as a mode-1 tensor. A $N$-mode, real-valued tensor is denoted by $\mathcal{X} \in \mathbb{R}^{I_0 \times \cdots \times I_{N-1}}$. This paper focuses on tensors of mode three or higher for tensor decomposition.

\subsubsection{Indices of a nonzero tensor element}~\label{index_intro}
For a 3-mode tensor, $\mathcal{X} \in \mathbb{R}^{I_0 \times I_1 \times I_2}$, a nonzero tensor element is indicated as $x = \mathcal{X}(i_0,i_1,i_2)$. Here, $i_0$, $i_1$, and $i_2$ are the positions or coordinates of $x$ in the tensor $\mathcal{X}$, which are commonly referred to as indices of the tensor element.

\subsubsection{Tensor matricization} $\mathcal{X}_{(n)}$ denotes the mode-$n$ matricization or matrix unfolding~\cite{favier2014overview} of $\mathcal{X}$. $\mathcal{X}_{(n)}$ is defined as the matrix $\mathcal{X}_{(n)} \in \mathbb{R}^{I_n \times (I_0 \cdots I_{n-1} I_{n+1} \cdots I_{N-1})}$ where the parenthetical ordering indicates, the mode-$n$ column vectors are arranged by sweeping all the other mode indices through their ranges.

\subsubsection{Canonical Poliyedic Tensor Decomposition (CPD)}
CPD decomposes $\mathcal{X}$ into a sum of single-mode tensors (i.e., arrays), which best approximates $\mathcal{X}$. For example, given 3-mode tensor $\mathcal{X} \in \mathbb{R}^{I_0 \times I_1 \times I_2}$, our goal is to approximate the original tensor as $\mathcal{X} \approx \sum_{r=0}^{R-1} \mathbf{a}_r \circ \mathbf{b}_r \circ \mathbf{c}_r$, where $R$ is a positive integer and $\mathbf{a}_r \in \mathbb{R}^{I_0}$, $\mathbf{b}_r \in \mathbb{R}^{I_1}$, and $\mathbf{c}_r \in \mathbb{R}^{I_2}$.



For each of the three modes, the spMTTKRP operation can be expressed as
\begin{equation} \label{eqn_spMTTKRP0}
\mathbf{\tilde{A}} =  \mathcal{X}_{(0)} ( \mathbf{{B}} \odot  \mathbf{{C}}), \text{ } 
\mathbf{\tilde{B}} = \mathcal{X}_{(1)} ( \mathbf{{C}} \odot  \mathbf{{A}}), \text{ }
\mathbf{\tilde{C}} = \mathcal{X}_{(2)}  ( \mathbf{{A}} \odot  \mathbf{{B}}) \text{ }
\end{equation}

The alternating least squares (ALS) method is used to compute CPD. In a 3-mode tensor, CPD sequentially performs the computations in Equation~\ref{eqn_spMTTKRP0}, iteratively. This can be generalized to higher mode tensors. Note that the matricization of $\mathcal{X}$ is different for each factor matrix computation. In this paper, performing MTTKRP on all the matricizations of an input tensor is called computing MTTKRP along all the modes. The outputs $\mathbf{A}$, $\mathbf{B}$, and $\mathbf{C}$ are the factor matrices that approximate $\mathcal{X}$. $\mathbf{a}_r$, $\mathbf{b}_r$, and $\mathbf{c}_r$ refers to the $r^{\text{th}}$ column of $\mathbf{A}$, $\mathbf{B}$, and $\mathbf{C}$, respectively.

In this paper, we focus on MTTKRP on sparse tensors, which means that the tensor is sparse. Note that the factor matrices are dense.

\subsection{Related Work}
FLYCOO-GPU~\cite{wijeratne2024sparse} proposes a single GPU-based parallel algorithm to accelerate MTTKRP on sparse tensors while adopting the FLYCOO tensor format~\cite{10.1145/3543622.3573179}. FLYCOO-GPU~\cite{10.1145/3543622.3573179} introduces dynamic tensor remapping on the GPU to reorder the tensor during execution time, allowing mode-specific optimizations. Dynamically remapping the input tensor limits the scalability of FLYCOO-GPU across GPUs as the inter-GPU communication becomes the bottleneck in multiple GPUs. Unlike~\cite{wijeratne2024sparse}, our work supports multiple GPUs while avoiding inter-GPU task dependencies that lead to race conditions across GPUs.

BLCO~\cite{10.1145/3524059.3532363} proposes the blocked linearized coordinate format (BLCO) that enables out-of-memory computation where the input tensor is stored in host CPU external memory and streamed to a single GPU during the execution time of each mode computation. The streaming data-based computations of BLCO enable the partitions of the input tensor to be stored inside a large host CPU external memory, which is loaded into a single GPU over time and performs the MTTKRP partition by partition. Unlike BLCO, our approach distributes the computations across several GPUs while balancing the workload among all the GPUs.

HPSPTM~\cite{9835348} proposes a framework to exploit multilevel parallelism and data reusability in heterogeneous HPC systems, including CPU + multi-GPU platforms. The data scheduling and partitioning scheme of HPSPTM dynamically transfers tensor partitions, intermediate results, and rows of factor matrices between GPUs during each factor matrix computation. In contrast, our work introduces a static load balancing scheme to minimize the data transfers across GPUs during each factor matrix computation. HPSPTM focuses on million-scale tensors, while our work focuses on billion-scale tensors.

MM-CSF~\cite{10.1145/3295500.3356216} and HiCOO-GPU~\cite{li2018parti} propose novel tensor formats to distribute workload on a single GPU. In contrast, we propose a parallel algorithm to perform MTTKRP using multiple GPUs, which requires optimizations to reduce communication and synchronization overheads across GPUs.

Table~\ref{tab:relatedwork} compares the characteristics of the related work with our work. Unlike related work, our work supports billion-scale sparse tensors. Our work also uses a task-independent partitioning scheme that balances the workload among the GPUs and their streaming multiprocessors.

\section{Tensor Partitioning Scheme}~\label{Tensor_Format}~\label{note_sec}~\vspace{-3mm}

In this work, we extend the partitioning scheme proposed in FLYCOO-GPU~\cite{wijeratne2024sparse} to distribute the nonzero tensor elements among multiple GPUs. Specifically, we adopt the FLYCOO-GPU partitioning scheme to distribute the tensor across GPUs, leveraging its task independence property, which allows each tensor partition to operate independently of the other partitions. For completeness, we describe the adopted partitioning scheme as inter-device partitioning.

Unlike FLYCOO-GPU~\cite{wijeratne2024sparse}, our approach does not use dynamic tensor remapping. Consequently, we avoid embedding shard IDs within each nonzero tensor element. Instead, we maintain multiple copies of the input tensor in CPU external memory. For a detailed explanation of dynamic tensor remapping and the use of shard IDs, please refer to~\cite{wijeratne2024sparse}.

Furthermore, we use the same notation introduced by FLYCOO-GPU~\cite{wijeratne2024sparse}: When performing MTTKRP for the mode $d$ of an input tensor, we denote the mode $d$ as the output mode and its corresponding factor matrix as the output factor matrix. The rest of the tensor modes are called input modes, and the corresponding factor matrices are called input factor matrices.

\subsubsection{Elementwise computation (EC)}~\label{row_update_computation}

\begin{figure}[ht]
  \begin{center}
    \includegraphics[width=0.4\textwidth]{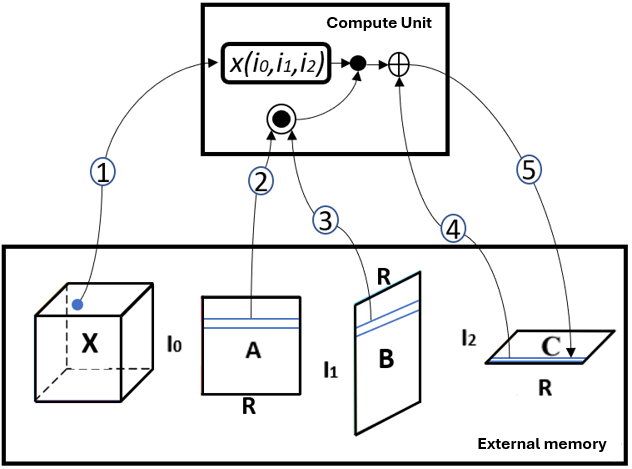}
  \end{center}
  \caption{Elementwise computation~\cite{wijeratne2024sparse}}
  \label{element_fig}
\end{figure}

Figure~\ref{element_fig} summarizes the elementwise computation of a nonzero tensor element in mode 2 of a 3-mode tensor.

In Figure~\ref{element_fig}, the elementwise computation is carried out on a nonzero tensor element, denoted as $\mathcal{X}_{(2)}(i_0,i_1,i_2)$. In sparse tensors, $\mathcal{X}_{(2)}(i_0,i_1,i_2)$ is typically represented in formats such as COOrdinate (COO) format. These formats store the indices ($i_0$, $i_1$, and $i_2$) along with the element value (i.e., $val(\mathcal{X}_{(2)}(i_0,i_1,i_2))$).

To perform the computation, $\mathcal{X}_{(2)}(i_0,i_1,i_2)$ is first loaded onto the processing units (i.e., streaming multiprocessors for GPU) from the external memory (step \circled{1}). The compute device retrieves the rows $\mathbf{{A}}(i_0,:)$, $\mathbf{{B}}(i_1,:)$, and $\mathbf{{C}}(i_2,:)$ from the factor matrices using the index values extracted from $\mathcal{X}_{(2)}(i_0,i_1,i_2)$ (step \circled{2}, step \circled{3}, and step \circled{4}). Then, the compute device performs the following computation:
\[
\mathbf{{C}}(i_2,r) = \mathbf{{C}}(i_2,r) + val(\mathcal{X}_{(2)}(i_0,i_1,i_2)) \cdot \mathbf{{A}}(i_0,r) \cdot \mathbf{{B}}(i_1,r)
\]
Here, $r$ refers to the column index of a factor matrix row ($r < R$). The operation involves performing a Hadamard product between row $\mathbf{{A}}(i_0,:)$ and row $\mathbf{{B}}(i_1,:)$, and then multiplying each element of the resulting product by $val(\mathcal{X}_{(2)}(i_0,i_1,i_2))$. Finally, the updated value is stored in the external memory (step \circled{5}).

\subsection{Tensor Partitioning}~\label{Tensor_Format_Definition}
We introduce a static partitioning scheme to partition an input tensor along each mode of the input tensor. Similarly to related works~\cite{8821030, 10.1145/3295500.3356216}, we use multiple copies of the input tensor, where each tensor copy is partitioned targeting an output mode computation. Note that all tensor copies are stored in the large external memory of the host CPU. 

Our partitioning scheme in each mode distributes the nonzero tensor elements across (1) multiple GPUs using tensor sharding and (2) streaming multiprocessors (SMs) within each GPU using inter-shard partitioning. Note that a tensor shard is executed by a single GPU, and an inter-shard partition is executed by a single GPU SM.

\subsubsection{Tensor Sharding}
The proposed sharding scheme organizes the nonzero tensor elements of the input tensor into \textit{tensor shards} (\textit{TS}) based on the output mode index along each mode. For a given output mode $d$, the EC introduced in Section~\ref{row_update_computation} is performed on each nonzero tensor element to update the corresponding rows of the output factor matrix. All nonzero tensor elements sharing the same output factor matrix index contribute to updating the same row. GPUs should maintain coherence among the updates that use nonzero tensor elements with the same output mode indices to avoid race conditions. To prevent race conditions between GPUs, all nonzero tensor elements sharing the same output mode index are assigned to the same \textit{TS}, eliminating task dependencies across GPUs. It avoids the need to maintain coherency across GPUs during execution, thereby significantly reducing synchronization overhead.
\subsubsection{Inter-Shard Partitioning}
The nonzero tensor elements of each \textit{tensor shard} (\textit{TS}) are equally distributed among the SMs of each GPU by partitioning each \textit{TS} into equal-sized \textit{inter-shard partitions} (\textit{ISP}). With \textit{ISP} partitions, all the SMs of a GPU are assigned the same workload during execution. We use atomic operations to avoid race conditions between SMs of the same GPU.

\subsection{Tensor Format Definition}~\label{Tensor_Format_Definition_0}
Consider a multi-GPU setup with $m$ GPUs, each containing $g$ GPU SMs. For each output mode $d$, we divide the output mode indices $I_d$ into sets of equal-sized partitions $I_{d,0}, I_{d,1}, \ldots, I_{d, k_d - 1}$, where $k_d = \frac{|I_d|}{m}$. Here, $|I_d|$ denotes the size of $I_d$. Each index partition $I_{d, j}$ ($j = 0, 1, \ldots, (k_d -1)$) is a subset of the output mode indices $I_d$. Next, all nonzero tensor elements that incident on the indices in $I_{d, j}$ are collected into a tensor shard, denoted by $\textit{TS}_{d,j}$. 

To distribute nonzero tensor elements among GPU SMs, we further divide each tensor shard into equal-sized sets called inter-shard partitions (\textit{ISP}). Each tensor shard $\textit{TS}_{d,j}$ is divided into $t_{d,j} = \left\lceil|\textit{TS}_{d,j}|/g\right\rceil$ inter-shard partitions. We denote the $q$-th inter-shard partition in $\textit{TS}_{d,j}$ as $\textit{ISP}_{d,j,q}$. For a tensor with $|\mathcal{T}|$ nonzero tensor elements, the total number of inter-shard partitions in mode $d$ is $\tau_d = \sum_{h=0}^{k_d-1} t_{d,h} \approx \frac{|\mathcal{T}|}{(m \times g)}$.

\section{Parallel Algorithm}\label{secparallel_algo}

\subsection{CUDA Programming Model}
In the CUDA programming model~\cite{4490127, 10.1145/3570638}, a multi-threaded program is partitioned into blocks of threads (i.e., threadblocks) where each threadblock operates independently. The threadblocks are organized into a GPU Grid~\cite{cuda2021cuda, ansorge2022programming}. A multi-GPU implementation executes a kernel as Grids of threadblocks where each Grid is executed on a separate GPU. Each threadblock is executed by one GPU streaming multiprocessor (SM).

\subsection{Tensor Partition and Threadblock Mapping}~\label{TensorPartitiontoThreadblockMapping}

\begin{figure}[ht]
\vspace{-3mm}
\centering
\includegraphics[width=0.85\linewidth]{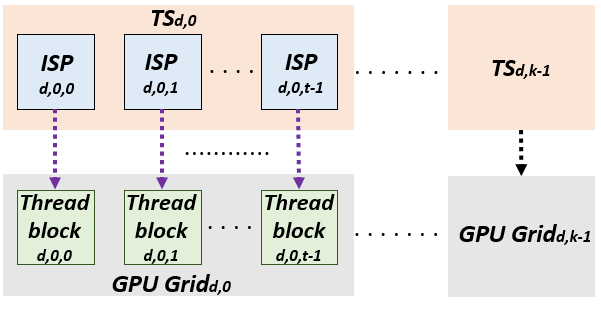}
\caption{Tensor partition mapping of mode \textit{d}}
\label{mapping_threads}
\vspace{-3mm}
\end{figure}

Following the partition scheme proposed in Section~\ref{Tensor_Format_Definition}, the tensor shards (\textit{TS}) and the inter-shard partitions (\textit{ISP}) are assigned to GPU Grids and threadblocks, respectively~\cite{cuda2021cuda}, as illustrated in Figure~\ref{mapping_threads}. Figure~\ref{mapping_threads} uses the same notation introduced in Section~\ref{Tensor_Format_Definition}. Once a GPU finishes executing all the computations in a Grid, a new Grid is loaded onto the GPU for execution. Similarly, when a GPU SM finishes executing all the computations in a threadblock, a new threadblock from the same Grid is assigned to the SM.

\subsection{Target Platform}~\label{Target_Platform}
\begin{figure}[ht]
\vspace{-4mm}
\centering
\includegraphics[width=0.9\linewidth]{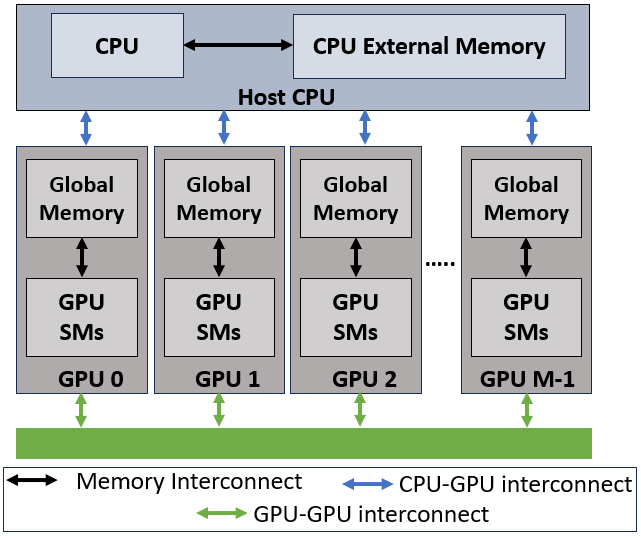}
\caption{Target platform}
\label{highlevel}
\end{figure}

We consider a single node multi-GPU platform where the GPUs are connected to a single CPU, as illustrated in Figure~\ref{highlevel}. All the GPUs are connected using GPU to GPU interconnection (i.e., GPUDirect P2P~\cite{nvidia_gpudirect}). Using the Cuda programming model, our parallel algorithm directly transfers data from one GPU global memory to another GPU global memory when required (see Section~\ref {secparallel_algo}).

In Figure~\ref{highlevel}, we show GPU to GPU interconnection and host CPU to GPU interconnection separately for clarity. The host CPU to GPU interconnection and GPU to GPU interconnection can share the same physical hardware connections such as PCIe.

\subsection{Data Distribution Among GPUs}~\label{data_distribution}
All the tensor shards (for all output modes) are stored in the host CPU external memory. When a tensor shard ($TS$) is ready for execution, it is transferred to the global memory of the corresponding GPU.

The factor matrices of each billion-scale tensor are significantly smaller (i.e., a few megabytes) than the GPU global memory. Each GPU maintains a local copy of the factor matrices in its global memory. Once the MTTKRP computation for an output mode is completed, the updated rows of the output factor matrix are exchanged between GPUs to prepare for the computation of the next output mode.

\subsection{Overall Algorithm}~\label{OverallAlgorithmMTTKRP}
In this Section, we present the parallel algorithm for a single iteration of tensor decomposition. In tensor decomposition, the proposed algorithm is iteratively performed to generate the factor matrices that best approximate the original input tensor.

\subsection{Mode-by-mode MTTKRP}

Algorithm~\ref{parallel_alg} shows the parallel algorithm for performing MTTKRP. Algorithm~\ref{parallel_alg} takes (1) all the input tensor copies $\textbf{T}$ and (2) factor matrices denoted as $\textbf{Y} = \{Y_0, Y_1,..., Y_{N-1}\}$. As shown in Algorithm~\ref{parallel_alg}, the MTTKRP is performed mode by mode (Algorithm~\ref{parallel_alg}: line 6). In each mode, A GPU grid (Algorithm~\ref{parallel_alg}: line 7) operates on a tensor shard mapped onto a GPU. At the end of all the computations of one mode, the GPUs are globally synchronized, and the generated output factor matrix rows are exchanged across GPUs before the computations of the next mode to maintain the correctness of the program (Algorithm~\ref{parallel_alg}: lines 8 - 11).

\begin{algorithm}[ht]
    \DontPrintSemicolon
    Input: Input tensor copies, $\textbf{T} = \{T_0, T_1 \cdots T_{N-1}\}$ \;
    Randomly initialized factor matrices, $\textbf{Y} = \{Y_0, Y_1,...,Y_{N-1}\}$\;
    Output: Updated factor matrices $\hat{\textbf{Y}} = \{\hat{Y}_0, \hat{Y}_1,...,\hat{Y}_{N-1}\}$ \;
    \For{each mode $d = 0, \ldots, N-1$} {
    $T_{in} \leftarrow T_d$ \;
    \textcolor{blue}{// Execute Grids using multiple GPUs} \;
    \For{\text{each tensor shard}, $TS_{d,z}$ in $T_{in}$ \textbf{parallel}}{
    $\hat{Y}_d$ $\leftarrow$ \textbf{GPU Grid}($TS_{d,z}$, $\textbf{Y}$) \;
    }
    \textbf{\_\_Inter-GPU Barrier\_\_} \;
    \textcolor{blue}{// Exchange generated output factor matrix partition across GPU} \;
    $Y_d$ $\leftarrow$ \textbf{All Gather}($\hat{Y}_d$) \;
    \textbf{\_\_Inter-GPU Barrier\_\_} \;
    }
\caption{Overall Proposed Algorithm}
\label{parallel_alg}
\end{algorithm}

After processing $TS$, the updated factor matrix rows are shared across the GPUs (Algorithm~\ref{parallel_alg}: line 10) using all-gather communication primitive~\cite{kobus2019gossip} before executing MTTKRP for the next output mode.

\subsection{Mapping Parallel Algorithm to GPU threadblocks}~\label{sec_grid_model}~\label{sec_proposed_design}
\begin{figure}[ht]
\vspace{-5mm}
\centering
\includegraphics[width=0.8\linewidth]{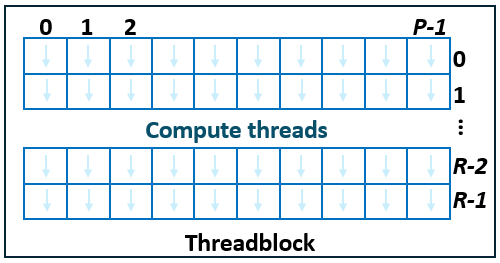}
\caption{Overview of threadblock}
\label{threadblock_over}
\end{figure}

\begin{algorithm}[ht]
    \DontPrintSemicolon
\textbf{GPU Grid}($TS_{d,z}$, $\textbf{Y}$, $T_{out}$):{ \;
    \textbf{Input}: Input tensor shard, $TS_{d,z}$\; Factor matrices $\textbf{Y} = \{Y_0, Y_1,...,Y_{N-1}\}$\;

    \textbf{Output}: Updated factor matrix of mode $d$, $\hat{Y}_d$\;

    \textcolor{blue}{// Execute threadblocks in parallel across GPU SMs} \;
     \For{each inter-shard partition $ISP_{d,z,b}$ in $TS_{d,z}$, $b = 0, 1, \ldots, t_{d,z-1}$ \textbf{parallel}}{
    \For{each column, $t$ in threadblock \textbf{parallel}}{
        \If{$nnz + t < |ISP_{d,z}|$}{
        \textbf{Load}($x_i$ at $(nnz+t)$) \;
            $value \leftarrow val_i$ \;
            $p_i = (c_0, \ldots, c_{N-1})$ \;
            \textcolor{blue}{// Elementwise Computation} \;
    \For{input mode $w\in\{0,\ldots,N-1\}\setminus\{d\}$}{

        $vec \leftarrow $ \textbf{Load}(row $c_w$ from $w^\text{th}$ factor matrix) \;
        \textcolor{blue}{// Row $0$ to $R-1$ of the threadblock perform independent computations} \;
        \For{each rank $r$ in $R$ \textbf{parallel}}{
            $\ell(r) \leftarrow \ell(r) \times vec(r)$ \;
        }
    }
    }
        \For{each rank $r$ in $R$ \textbf{parallel}}{
        $\hat{Y}_d(c_d, r) \leftarrow \text{Atomic}(\hat{Y}_d(c_d, r) + \ell(r))$
    }
    }
     \textcolor{blue}{// P is the number of columns in a threadblock} \;
    $nnz \leftarrow nnz + P$ \;
    }
}

\caption{Parallel Algorithm Executed on a Shard}
\label{parallel_alg_grid_op}
\end{algorithm}

The basic computing unit of a GPU is a thread. According to the GPU programming model~\cite{cuda2021cuda, ansorge2022programming}, a multi-threaded program is partitioned into blocks of threads (i.e., threadblocks) that operate independently.

In our proposed algorithm, a threadblock has a dimension of $R \times P$, where $R$ denotes the rank of factor matrices and $P$ indicates the number of nonzero tensor elements parallelly loaded to a threadblock (see Figure~\ref{threadblock_over}). Here, each column of the threadblock shares the same nonzero tensor element, and each column performs elementwise computation on a nonzero tensor element.

Algorithm~\ref{parallel_alg_grid_op} outlines the computations executed on a shard partition. In Algorithm~\ref{parallel_alg_grid_op}, $ISP_{d,z,p}$ corresponds to $p^{\text{th}}$ inter-shard partition in $z^{\text{th}}$ tensor partition of mode $d$. When a GPU SM is idle, a threadblock and its corresponding inter-shard partition are assigned to the GPU SM for computation. Each column in the threadblock loads a single nonzero tensor element at a time and shares the nonzero tensor element across the threads in the same column. Each thread in a column extracts the information from the tensor element $x_i$ in COO format (Algorithm~\ref{parallel_alg_grid_op}: lines 9-11). Subsequently, each threadblock performs elementwise computation (Algorithm~\ref{parallel_alg_grid_op}: lines 6-17). To achieve threadwise parallelism, each thread in a column only executes the update operation on a single column of a row of the output factor matrix (Algorithm~\ref{parallel_alg_grid_op}: lines 15 - 17). The rows of the input factor matrices are loaded from the GPU global memory (Algorithm~\ref{parallel_alg_grid_op}: lines 13-14) depending on the indices of the current tensor element ($p_i$) executed in the GPU thread. Each GPU threadblock locally updates the output factor matrix (Algorithm~\ref{parallel_alg_grid_op}: lines 15) while maintaining the coherence of each threadblock to ensure the correctness of the program. According to the proposed partitioning scheme, threadblocks require atomic operations for tensor elements that reside in the same tensor shard. Hence, we use atomic operations across the GPU threadblocks within the same GPU to maintain the correctness of the program (Algorithm~\ref{parallel_alg_grid_op}: lines 18-19).

\subsection{Host CPU-GPU Communication}~\label{cpu-gpu_communication}
The input tensor copies are initially stored in the host CPU memory. The tensor shards are transferred from the host CPU memory to each GPU global memory during the execution time of each output mode.

\subsection{All Gather Communication (GPU to GPU)}~\label{gpu-gpu_communication}
All GPUs maintain a local copy of the factor matrices in GPU global memory (see Section~\ref{data_distribution}). As mentioned in Section~\ref{data_distribution}, the factor matrices of each billion-scale tensor are significantly smaller (i.e., a few megabytes) than the GPU global memory. Hence, keeping a local copy of the factor matrices does not consume significant additional memory on each GPU. 

For a given output mode, once a GPU completes processing the \textit{tensor shard} (\textit{TS}) assigned to it, the GPU communicates the updated rows to all the GPUs in the platform. The distributed factor matrix will act as an input factor matrix in the following modes of MTTKRP computations.

We adopt the ring network communication model~\cite{8573483} to perform all-gather among the GPUs, as shown in Algorithm~\ref{alg_p2p_comm}. We use GPU-to-GPU communication to send and receive partitions of the factor matrices across GPUs without involving the CPU host memory. The ring network model is suitable for bulk transfers among neighboring devices with limited bandwidth, which is ideal for communicating the output factor matrix among the GPUs.

\begin{algorithm}[ht]
    \DontPrintSemicolon
\textbf{All Gather ($\hat{Y}_{d,gpu\_id}$)}: \;
Input: Mode $d$ output factor matrix partition in the local GPU, $\hat{Y}_{d,gpu\_id}$\;
    \textcolor{blue}{// Using ring network communication model} \;
     \textcolor{blue}{//number of GPUs = $M$} \;
    \For{each GPU $z = 0, \ldots, (M-2)$}{
    \textcolor{blue}{// Send and Receive commands are executed in parallel} \;
    Send\_Copy = $\hat{Y}_{d,(gpu\_id+z) mod M}$\;
    Send(Send\_Copy, $(gpu\_id+1) mod M$, size(Send\_Copy)) \;\;
    Rec\_Copy = $\hat{Y}_{d,(gpu\_id-z-1) mod M}$ \;
    Receive(Rec\_Copy, $(gpu\_id-1) mod M$, size(Rec\_Copy)))\;
    Barrier \;
    }
\caption{GPU to GPU peer-to-peer communication}
\label{alg_p2p_comm}
\end{algorithm}

\section{Experimental Results}~\label{experiments}\vspace{-6mm}
\subsection{Experimental Setup}~\label{ex_setup}\vspace{-5mm}
\subsubsection{Platform}\label{sec_platform}


\begin{figure*}[ht]
\centering
\includegraphics[width=0.85\linewidth]{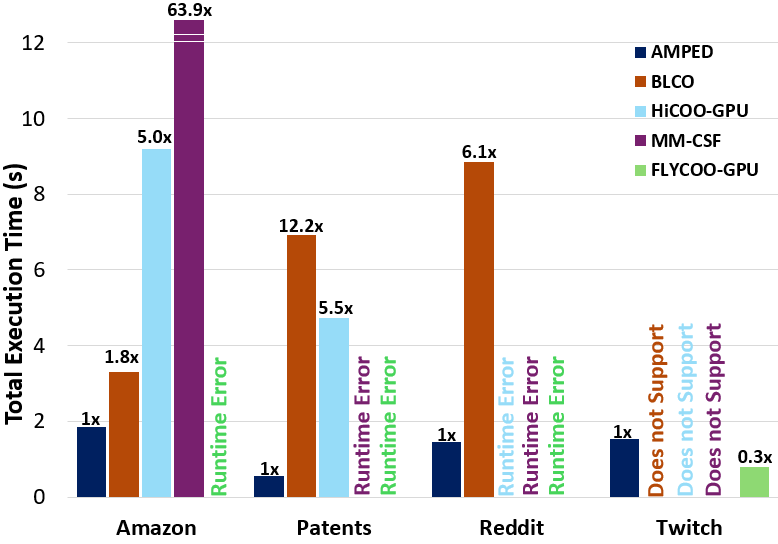}
\caption{Total execution time (speedup of our work compared to each baseline is shown at the top of each bar)}
\label{R32_TE}
\vspace{-2mm}
\end{figure*}

We conducted our experiments on a single CPU node with multi-GPUs. The CPU node has 4 NVIDIA RTX 6000 Ada Generation GPUs and AMD EPYC 9654 host CPU. The GPUs are connected to the host CPU via PCIe. GPUs use GPUDirect Peer-to-Peer (P2P)~\cite{nvidia_gpudirect} to communicate with each other. Note that NVIDIA RTX 6000 Ada GPUs do not support NVLink~\cite{nvidia_NVLINK} for direct GPU communication. Hence, we use the Direct P2P communication in our experiments.\\
\textit{GPU Specification:} NVIDIA RTX 6000 Ada Generation GPU features the Ada Lovelace GPU architecture with 142 Streaming Multiprocessors (SMs) and 18176 cores, sharing 48 GB of GDDR6 global memory. Note that the NVIDIA RTX 6000 Ada Generation GPU has more computing power and global memory than the NVIDIA A100 GPU.\\
\textit{Host CPU Specification:} In our experiments, we use a 2-socket AMD EPYC 9654 CPU as the host CPU platform. Each AMD EPYC 9654 consists of 96 physical cores (192 threads) running at a frequency of 2.4 GHz, sharing 1.5 TB of CPU external memory.

Each GPU is connected to the host CPU through a PCIe interface with 64 GB/s data bandwidth.


\begin{table}[ht]
\vspace{-3.5mm}
\caption{Characteristics of the sparse tensors}
\begin{center}
\resizebox{\columnwidth}{!}{
\begingroup
\setlength{\tabcolsep}{1.0pt} 
\renewcommand{\arraystretch}{1.5} 
\scriptsize 
\begin{tabular}{|c|c|c|}
 \hline
 \textbf{Tensor} & \textbf{Shape} & \textbf{Number of} \\
 & & \textbf{Tensor Elements} \\
 \hline\hline
 Amazon~\cite{frosttdataset} & $4.8M \times 1.8M \times 1.8M$ & $1.7B$ \\
 \hline
 Patents~\cite{frosttdataset} & $46 \times 239.2K \times 239.2K$ & $3.6B$ \\
 \hline
 Reddit-2015~\cite{frosttdataset} & $8.2M \times 177K \times 8.1M$ & $4.7B$ \\
 \hline
 Twitch~\cite{10.1145/3460231.3474267, recommend_repo} & $15.5M \times 6.2M \times 783.9K \times 6.1K \times 6.1K$ & $0.5B$ \\
 \hline
\end{tabular}
\endgroup
}
\label{table3}
\end{center}
\vspace{-7mm}
\end{table}

\subsubsection{Implementation}
The source code is developed using CUDA C++~\cite{ansorge2022programming} and compiled using CUDA version 12.2~\cite{fatica2008cuda}.

\subsubsection{Datasets}
We used all publicly available billion-scale tensors from the Formidable Repository of Open Sparse Tensors and Tools (FROSTT) dataset~\cite{frosttdataset} and Recommender Systems and Personalization Datasets~\cite{10.1145/3460231.3474267, recommend_repo}. Table~\ref{table3} summarizes the characteristics of the tensors.

\subsubsection{Baselines}~\label{baselines_exp}
We evaluate the performance of our work by comparing it with the state-of-the-art GPU implementations, BLCO~\cite{10.1145/3524059.3532363} (out-of-memory computation enabled), MM-CSF~\cite{8821030}, HiCOO-GPU \cite{li2018parti}, and FLYCOO-GPU~\cite{wijeratne2024sparse}. The out-of-memory computation of BLCO stores the partitions of the input tensor inside the host CPU external memory, which is loaded into a single GPU over time and performs the MTTKRP partition by partition. To achieve optimal results with HiCOO-GPU, we use the recommended configurations provided in the source code~\cite{hicoo_repo}. For our experiments, we utilize the open-source BLCO repository~\cite{blco_github}, ParTI repository~\cite{hicoo_repo}, and MM-CSF~\cite{mm_csf} repository.





\subsubsection{Default Configuration}~\label{standard_config}
In our experiments, we used 4 GPUs as the default number of GPUs in the system. We set $\theta = 32$ and $R = 32$ as the configuration of the threadblocks described in Section~\ref{OverallAlgorithmMTTKRP}.

\subsubsection{Performance Metric - Total Execution Time} Similar to the literature~\cite{8821030, wijeratne2024sparse, 10.1145/3524059.3532363, li2018parti}, we measure the performance using the execution time to compute MTTKRP across all modes of an input tensor in a single iteration of tensor decomposition.

\subsection{Overall Performance}~\label{overall_perf_exp}
Figure~\ref{R32_TE} shows the total execution time of our work on 4 NVIDIA RTX 6000 Ada GPUs. The speedup achieved by our work compared with each baseline in each input tensor is displayed at the top of the respective bar. The runtime error indicates that the host CPU operating system terminated the baseline during execution due to insufficient memory on the target hardware to store the input tensor, factor matrices and intermediate values.

For evaluation, we set the rank of the factor matrices ($R$) to 32, similar to the state-of-the-art~\cite{8821030, 10.1145/3524059.3532363, li2018parti}. Our work demonstrates a geometric mean speedup of 5.1$\times$ compared with the state-of-the-art baselines.

When out-of-memory computation is enabled, BLCO~\cite{10.1145/3524059.3532363} stores the input tensor in the host CPU memory and loads the tensor partition by partition from the external memory of the host CPU to the GPU global memory. In BLCO, using a single GPU introduces additional memory traffic between the host CPU and the GPU. Meanwhile, our work has more effective bandwidth between the host CPU and the GPU since multiple GPUs can concurrently communicate with the host CPU. Our work shows a geometric speedup of $5.1\times$ using 4 GPUs compared to BLCO.

MM-CSF~\cite{8821030} performs MTTKRP only on the Amazon dataset. For  Patents and Reddit, the GPU ran out of memory during the execution time. ParTI-GPU~\cite{li2018parti} can perform MTTKRP on Amazon and Patents. Also, MM-CSF~\cite{8821030} and ParTI-GPU~\cite{li2018parti} do not support Twitch, which has 5 modes.

FLYCOO-GPU~\cite{wijeratne2024sparse} does not support Amazon, Patents, and Reddit since these tensors do not fit in the GPU global memory. FLYCOO-GPU requires maintaining 2 copies of the tensor in the GPU global memory. On Twitch, FLYCOO-GPU outperforms our work by 3.9$\times$ due to the communication overhead of our work. Twitch is the smallest billion-scale tensor in the literature. The small size enables keeping 2 tensor copies inside the global memory of a single GPU. Since FLYCOO-GPU only targets a single GPU, FLYCOO-GPU does not require GPU-GPU communication or host CPU-GPU communication.



\subsection{Impact of Partitioning Scheme}~\label{scalability}
\begin{figure}[ht]
\vspace{-7mm}
  \begin{center}
    \includegraphics[width=0.45\textwidth]{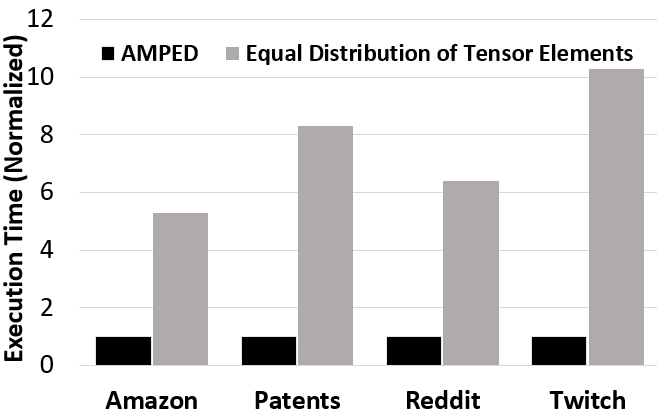}
  \end{center}
  \caption{Impact of proposed partitioning scheme}
  \label{impact_ss_schedule}
  \vspace{-2mm}
\end{figure}

Section~\ref{Tensor_Format} describes the partitioning scheme used in this work to distribute the tensor shards among GPUs.

An alternative approach is to distribute the non-zero tensor elements equally among all GPUs. It introduces additional computations on the host CPU to merge the partial results of each tensor shard. Figure~\ref{impact_ss_schedule} compares our proposed strategy with the equal distribution of nonzero tensor elements among GPUs. Our proposed partitioning scheme achieves 5.3$\times$ to 10.3$\times$ speedups in total execution time.

\subsection{Execution Time Breakdown}~\label{ETB}

For large tensors such as Patents and Reddit, moving tensor shards from CPU host memory to GPU global memory is the major contributor to the total communication time. Tensors with a large number of indices (e.g., Amazon and Twitch) require frequent GPU-GPU communication to update factor matrices at the end of each output mode computation, which significantly contributes to the total communication time. In particular, Reddit exhibits a significant communication overhead ($32\%$) during the execution due to (1) the large number of nonzero tensor elements in the tensor and (2) the modes with a large number of indices.

\begin{figure}[ht]
  \begin{center}
    \includegraphics[width=0.45\textwidth]{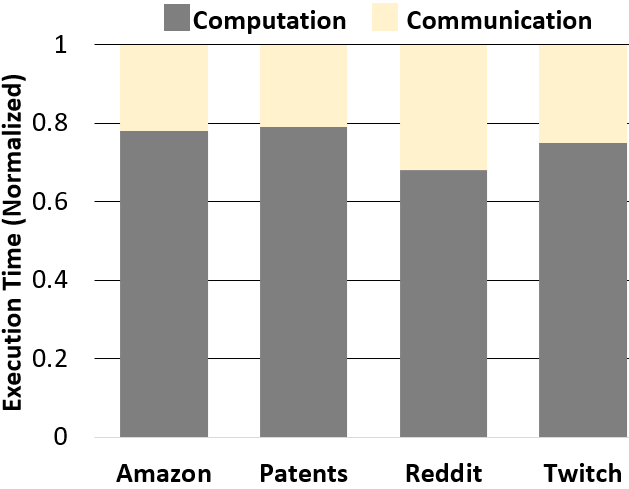}
  \end{center}
  \caption{Execution time breakdown of the input tensors}
  \label{ET_breakdown_mode}
  \vspace{-2mm}
\end{figure}


\subsection{Workload Distribution among GPUs}~\label{idle_time}
\begin{figure}[ht]
\vspace{-3mm}
  \begin{center}
    \includegraphics[width=0.45\textwidth]{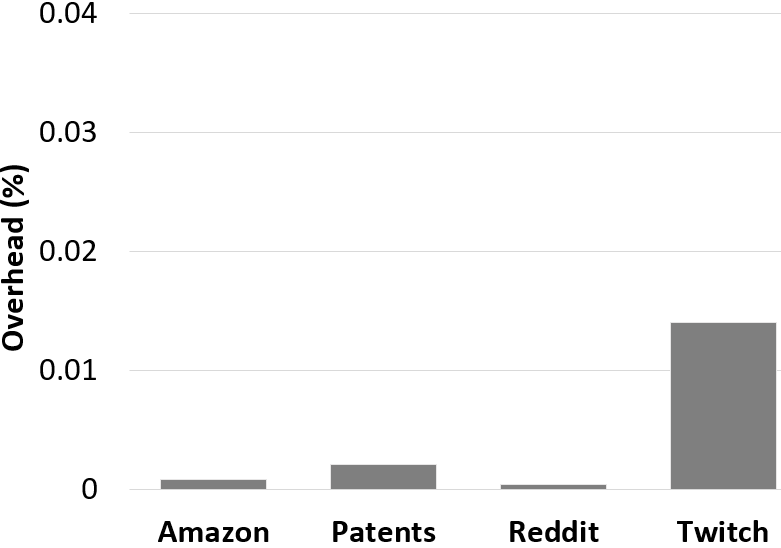}
  \end{center}
  \caption{Computation time overhead among GPUs}
  \label{overhead_fig}
  \vspace{-4mm}
\end{figure}

As discussed in Section~\ref{secparallel_algo}, each GPU updates a set of rows of the output factor matrix. In this section, the computation time of a GPU is defined as the total time to perform elementwise computation (EC) on all nonzero tensor elements assigned to a GPU across all modes. The computation time overhead is the difference between the maximum and minimum computation times among GPUs on the target platform.

Figure~\ref{overhead_fig} shows the computation time overhead among GPUs (as a percentage) on our target 4 GPU platform. To determine the computation time of each GPU, we execute each GPU grid (see Section~\ref{TensorPartitiontoThreadblockMapping}) separately and measure the computation time in each output mode. The computation time overhead for all billion-scale tensors is less than $1\%$ (as a percentage of the total time required to perform all EC using all 4 GPUs).

However, Twitch has the most computation time overhead due to some indices of the tensor corresponding to popular streamers and games in the Twitch platform, which leads to a disproportionately large number of nonzero tensor elements assigned to those indices. It results in a workload imbalance between GPUs, leading to a larger overhead.

\subsection{Scalability}~\label{L1-util}
\begin{figure}[ht]
  \begin{center}
    \includegraphics[width=0.45\textwidth]{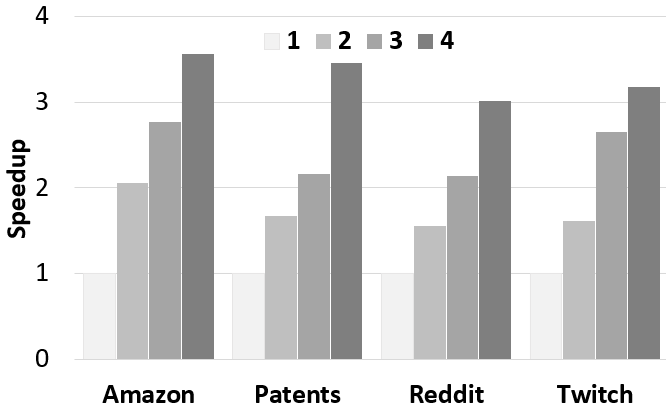}
  \end{center}
  \caption{Scalability of the proposed Algorithm}
  \label{Scalability_fig}
  \vspace{-3mm}
\end{figure}

Figure~\ref{Scalability_fig} shows the speedup of each input tensor as the number of GPUs increases from 1 to 4. The speedup is calculated by dividing the total execution time of the input tensor on a single GPU by its total execution time in each case. Our proposed parallel algorithm achieves geometric mean speedups of 1.9$\times$, 2.3$\times$, and 3.3$\times$ when using 2, 3, and 4 GPUs, respectively, compared to a total execution time of a single GPU. The speedup increases nearly linearly with the number of GPUs for each tensor.

In the single GPU implementation, tensor partitions are stored in the host CPU external memory and loaded onto the GPU one \textit{iDp} at a time in each mode.

\subsection{Preprocessing Time}

\begin{figure}[ht]
\vspace{-2mm}
  \begin{center}
    \includegraphics[width=0.45\textwidth]{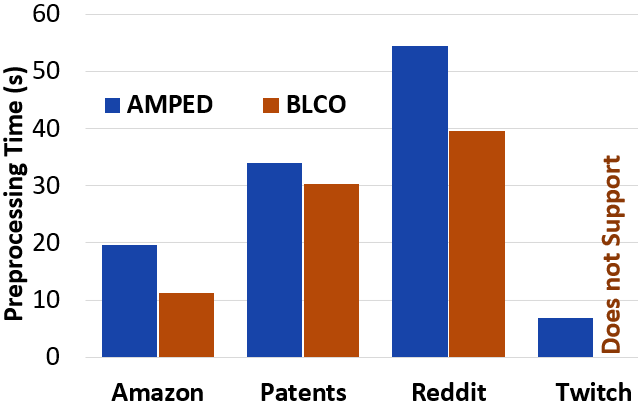}
  \end{center}
  \caption{Preprocessing time}
  \label{tensor_formation}
  \vspace{-2mm}
\end{figure}

The preprocessing of an input tensor involves generating tensor partitions following Section~\ref{Tensor_Format}. Note that our work does not focus on the acceleration of preprocessing time. We have included a comparison of preprocessing times in Figure~\ref{tensor_formation} for completeness. For comparison, we used the preprocessing time of BLCO. The host CPU described in Section~\ref{sec_platform} is used for preprocessing.

\section{Conclusion and Future Work}
In this paper, we proposed AMPED, a novel parallel algorithm designed to accelerate MTTKRP on billion-scale tensors using multiple GPUs. Our approach introduced a partitioning scheme that allowed each tensor partition to be executed independently of other partitions in each output mode. The partitioning scheme, coupled with the proposed load-balancing strategy to distribute the workload across all the GPUs, minimized the GPU idle time. AMPED achieved a geometric mean speedup of 5.1$\times$ (using 4 GPUs) in total execution time compared with the state-of-the-art GPU baselines.

In our future work, we will adapt the proposed parallel algorithm to heterogeneous computing platforms with different devices, such as multiple CPUs, GPUs, and Field Programmable Gate Arrays (FPGAs), to show the adaptability of the proposed parallel algorithm.

\begin{acks}
This work is supported by the National Science Foundation (NSF) under grant OAC-2209563, CSSI-2311870, and in part by the DEVCOM Army Research Lab under grant W911NF2220159.
\end{acks}

\bibliographystyle{ACM-Reference-Format}
\bibliography{chi_bib,ref}


\begin{thebibliography}{38}


\ifx \showCODEN    \undefined \def \showCODEN     #1{\unskip}     \fi
\ifx \showISBNx    \undefined \def \showISBNx     #1{\unskip}     \fi
\ifx \showISBNxiii \undefined \def \showISBNxiii  #1{\unskip}     \fi
\ifx \showISSN     \undefined \def \showISSN      #1{\unskip}     \fi
\ifx \showLCCN     \undefined \def \showLCCN      #1{\unskip}     \fi
\ifx \shownote     \undefined \def \shownote      #1{#1}          \fi
\ifx \showarticletitle \undefined \def \showarticletitle #1{#1}   \fi
\ifx \showURL      \undefined \def \showURL       {\relax}        \fi
\providecommand\bibfield[2]{#2}
\providecommand\bibinfo[2]{#2}
\providecommand\natexlab[1]{#1}
\providecommand\showeprint[2][]{arXiv:#2}

\bibitem[Ansorge(2022)]%
        {cuda2021cuda}
\bibfield{author}{\bibinfo{person}{Richard Ansorge}.} \bibinfo{year}{2022}\natexlab{}.
\newblock \bibinfo{booktitle}{\emph{Programming in parallel with CUDA: a practical guide}}.
\newblock \bibinfo{publisher}{Cambridge University Press}.
\newblock


\bibitem[Bharadwaj et~al\mbox{.}(2024)]%
        {10.1145/3626183.3659980}
\bibfield{author}{\bibinfo{person}{Vivek Bharadwaj}, \bibinfo{person}{Osman~Asif Malik}, \bibinfo{person}{Riley Murray}, \bibinfo{person}{Ayd\i{}n Bulu\c{c}}, {and} \bibinfo{person}{James Demmel}.} \bibinfo{year}{2024}\natexlab{}.
\newblock \showarticletitle{Distributed-Memory Randomized Algorithms for Sparse Tensor CP Decomposition}. In \bibinfo{booktitle}{\emph{Proceedings of the 36th ACM Symposium on Parallelism in Algorithms and Architectures}} (Nantes, France) \emph{(\bibinfo{series}{SPAA '24})}. \bibinfo{publisher}{Association for Computing Machinery}, \bibinfo{address}{New York, NY, USA}, \bibinfo{pages}{155–168}.
\newblock
\showISBNx{9798400704161}
\href{https://doi.org/10.1145/3626183.3659980}{doi:\nolinkurl{10.1145/3626183.3659980}}


\bibitem[Chen et~al\mbox{.}(2022a)]%
        {5318}
\bibfield{author}{\bibinfo{person}{Yuedan Chen}, \bibinfo{person}{Guoqing Xiao}, \bibinfo{person}{Tamer Ozsu}, \bibinfo{person}{Zhuo Tang}, \bibinfo{person}{Albert~Y. Zomaya}, {and} \bibinfo{person}{Kenli Li}.} \bibinfo{year}{2022}\natexlab{a}.
\newblock \showarticletitle{Exploiting Hierarchical Parallelism and Reusability in Tensor Kernel Processing on Heterogeneous HPC Systems}. In \bibinfo{booktitle}{\emph{IEEE International Conference on Data Engineering (ICDE)}}.
\newblock
\href{https://doi.org/10.1109/ICDE53745.2022.00234}{doi:\nolinkurl{10.1109/ICDE53745.2022.00234}}


\bibitem[Chen et~al\mbox{.}(2022b)]%
        {9835348}
\bibfield{author}{\bibinfo{person}{Yuedan Chen}, \bibinfo{person}{Guoqing Xiao}, \bibinfo{person}{M.~Tamer Özsu}, \bibinfo{person}{Zhuo Tang}, \bibinfo{person}{Albert~Y. Zomaya}, {and} \bibinfo{person}{Kenli Li}.} \bibinfo{year}{2022}\natexlab{b}.
\newblock \showarticletitle{Exploiting Hierarchical Parallelism and Reusability in Tensor Kernel Processing on Heterogeneous HPC Systems}. In \bibinfo{booktitle}{\emph{2022 IEEE 38th International Conference on Data Engineering (ICDE)}}. \bibinfo{pages}{2522--2535}.
\newblock
\href{https://doi.org/10.1109/ICDE53745.2022.00234}{doi:\nolinkurl{10.1109/ICDE53745.2022.00234}}


\bibitem[Cichocki(2014)]%
        {cichocki2014era}
\bibfield{author}{\bibinfo{person}{Andrzej Cichocki}.} \bibinfo{year}{2014}\natexlab{}.
\newblock \showarticletitle{Era of big data processing: A new approach via tensor networks and tensor decompositions}.
\newblock \bibinfo{journal}{\emph{arXiv preprint arXiv:1403.2048}} (\bibinfo{year}{2014}).
\newblock


\bibitem[Cichocki et~al\mbox{.}(2015)]%
        {cichocki2015tensor}
\bibfield{author}{\bibinfo{person}{Andrzej Cichocki}, \bibinfo{person}{Danilo Mandic}, \bibinfo{person}{Lieven De~Lathauwer}, \bibinfo{person}{Guoxu Zhou}, \bibinfo{person}{Qibin Zhao}, \bibinfo{person}{Cesar Caiafa}, {and} \bibinfo{person}{Huy~Anh Phan}.} \bibinfo{year}{2015}\natexlab{}.
\newblock \showarticletitle{Tensor decompositions for signal processing applications: From two-way to multiway component analysis}.
\newblock \bibinfo{journal}{\emph{IEEE signal processing magazine}} \bibinfo{volume}{32}, \bibinfo{number}{2} (\bibinfo{year}{2015}), \bibinfo{pages}{145--163}.
\newblock


\bibitem[Fatica(2008)]%
        {fatica2008cuda}
\bibfield{author}{\bibinfo{person}{Massimiliano Fatica}.} \bibinfo{year}{2008}\natexlab{}.
\newblock \showarticletitle{CUDA toolkit and libraries}. In \bibinfo{booktitle}{\emph{2008 IEEE hot chips 20 symposium (HCS)}}. IEEE, \bibinfo{pages}{1--22}.
\newblock


\bibitem[Favier and de~Almeida(2014)]%
        {favier2014overview}
\bibfield{author}{\bibinfo{person}{G{\'e}rard Favier} {and} \bibinfo{person}{Andr{\'e}~LF de Almeida}.} \bibinfo{year}{2014}\natexlab{}.
\newblock \showarticletitle{Overview of constrained PARAFAC models}.
\newblock \bibinfo{journal}{\emph{EURASIP Journal on Advances in Signal Processing}} \bibinfo{volume}{2014}, \bibinfo{number}{1} (\bibinfo{year}{2014}), \bibinfo{pages}{1--25}.
\newblock


\bibitem[Fernandes et~al\mbox{.}(2021)]%
        {fernandes2021tensor}
\bibfield{author}{\bibinfo{person}{Sofia Fernandes}, \bibinfo{person}{Hadi Fanaee-T}, {and} \bibinfo{person}{Jo{\~a}o Gama}.} \bibinfo{year}{2021}\natexlab{}.
\newblock \showarticletitle{Tensor decomposition for analysing time-evolving social networks: An overview}.
\newblock \bibinfo{journal}{\emph{Artificial Intelligence Review}} \bibinfo{volume}{54}, \bibinfo{number}{4} (\bibinfo{year}{2021}), \bibinfo{pages}{2891--2916}.
\newblock


\bibitem[Helal et~al\mbox{.}(2021)]%
        {alto_paper}
\bibfield{author}{\bibinfo{person}{Ahmed~E. Helal}, \bibinfo{person}{Jan Laukemann}, \bibinfo{person}{Fabio Checconi}, \bibinfo{person}{Jesmin~Jahan Tithi}, \bibinfo{person}{Teresa Ranadive}, \bibinfo{person}{Fabrizio Petrini}, {and} \bibinfo{person}{Jeewhan Choi}.} \bibinfo{year}{2021}\natexlab{}.
\newblock \showarticletitle{ALTO: Adaptive Linearized Storage of Sparse Tensors}. In \bibinfo{booktitle}{\emph{Proceedings of the ACM International Conference on Supercomputing}} (Virtual Event, USA) \emph{(\bibinfo{series}{ICS '21})}. \bibinfo{publisher}{Association for Computing Machinery}, \bibinfo{address}{New York, NY, USA}, \bibinfo{pages}{404–416}.
\newblock
\showISBNx{9781450383356}
\href{https://doi.org/10.1145/3447818.3461703}{doi:\nolinkurl{10.1145/3447818.3461703}}


\bibitem[Hijma et~al\mbox{.}(2023)]%
        {10.1145/3570638}
\bibfield{author}{\bibinfo{person}{Pieter Hijma}, \bibinfo{person}{Stijn Heldens}, \bibinfo{person}{Alessio Sclocco}, \bibinfo{person}{Ben van Werkhoven}, {and} \bibinfo{person}{Henri~E. Bal}.} \bibinfo{year}{2023}\natexlab{}.
\newblock \showarticletitle{Optimization Techniques for GPU Programming}.
\newblock \bibinfo{journal}{\emph{ACM Comput. Surv.}} \bibinfo{volume}{55}, \bibinfo{number}{11}, Article \bibinfo{articleno}{239} (\bibinfo{date}{March} \bibinfo{year}{2023}), \bibinfo{numpages}{81}~pages.
\newblock
\showISSN{0360-0300}
\href{https://doi.org/10.1145/3570638}{doi:\nolinkurl{10.1145/3570638}}


\bibitem[Hong et~al\mbox{.}(2020)]%
        {hong2020generalized}
\bibfield{author}{\bibinfo{person}{David Hong}, \bibinfo{person}{Tamara~G Kolda}, {and} \bibinfo{person}{Jed~A Duersch}.} \bibinfo{year}{2020}\natexlab{}.
\newblock \showarticletitle{Generalized canonical polyadic tensor decomposition}.
\newblock \bibinfo{journal}{\emph{SIAM Rev.}} \bibinfo{volume}{62}, \bibinfo{number}{1} (\bibinfo{year}{2020}), \bibinfo{pages}{133--163}.
\newblock


\bibitem[Ji et~al\mbox{.}(2019)]%
        {8884203}
\bibfield{author}{\bibinfo{person}{Yuwang Ji}, \bibinfo{person}{Qiang Wang}, \bibinfo{person}{Xuan Li}, {and} \bibinfo{person}{Jie Liu}.} \bibinfo{year}{2019}\natexlab{}.
\newblock \showarticletitle{A Survey on Tensor Techniques and Applications in Machine Learning}.
\newblock \bibinfo{journal}{\emph{IEEE Access}}  \bibinfo{volume}{7} (\bibinfo{year}{2019}), \bibinfo{pages}{162950--162990}.
\newblock
\href{https://doi.org/10.1109/ACCESS.2019.2949814}{doi:\nolinkurl{10.1109/ACCESS.2019.2949814}}


\bibitem[Kobus et~al\mbox{.}(2019)]%
        {kobus2019gossip}
\bibfield{author}{\bibinfo{person}{Robin Kobus}, \bibinfo{person}{Daniel J{\"u}nger}, \bibinfo{person}{Christian Hundt}, {and} \bibinfo{person}{Bertil Schmidt}.} \bibinfo{year}{2019}\natexlab{}.
\newblock \showarticletitle{Gossip: Efficient communication primitives for multi-gpu systems}. In \bibinfo{booktitle}{\emph{Proceedings of the 48th International Conference on Parallel Processing}}. \bibinfo{pages}{1--10}.
\newblock


\bibitem[Kolda and Bader(2009)]%
        {kolda2009tensor}
\bibfield{author}{\bibinfo{person}{Tamara~G Kolda} {and} \bibinfo{person}{Brett~W Bader}.} \bibinfo{year}{2009}\natexlab{}.
\newblock \showarticletitle{Tensor decompositions and applications}.
\newblock \bibinfo{journal}{\emph{SIAM review}} \bibinfo{volume}{51}, \bibinfo{number}{3} (\bibinfo{year}{2009}), \bibinfo{pages}{455--500}.
\newblock


\bibitem[Kurt et~al\mbox{.}(2022)]%
        {9820702}
\bibfield{author}{\bibinfo{person}{Süreyya~Emre Kurt}, \bibinfo{person}{Saurabh Raje}, \bibinfo{person}{Aravind Sukumaran-Rajam}, {and} \bibinfo{person}{P. Sadayappan}.} \bibinfo{year}{2022}\natexlab{}.
\newblock \showarticletitle{Sparsity-Aware Tensor Decomposition}. In \bibinfo{booktitle}{\emph{2022 IEEE International Parallel and Distributed Processing Symposium (IPDPS)}}. \bibinfo{pages}{952--962}.
\newblock
\href{https://doi.org/10.1109/IPDPS53621.2022.00097}{doi:\nolinkurl{10.1109/IPDPS53621.2022.00097}}


\bibitem[Laukemann et~al\mbox{.}(2024)]%
        {laukemann2024accelerating}
\bibfield{author}{\bibinfo{person}{Jan Laukemann}, \bibinfo{person}{Ahmed~E Helal}, \bibinfo{person}{S Anderson}, \bibinfo{person}{Fabio Checconi}, \bibinfo{person}{Yongseok Soh}, \bibinfo{person}{Jesmin~Jahan Tithi}, \bibinfo{person}{Teresa Ranadive}, \bibinfo{person}{Brian~J Gravelle}, \bibinfo{person}{Fabrizio Petrini}, {and} \bibinfo{person}{Jee Choi}.} \bibinfo{year}{2024}\natexlab{}.
\newblock \showarticletitle{Accelerating Sparse Tensor Decomposition Using Adaptive Linearized Representation}.
\newblock \bibinfo{journal}{\emph{arXiv preprint arXiv:2403.06348}} (\bibinfo{year}{2024}).
\newblock


\bibitem[Li et~al\mbox{.}(2018b)]%
        {8573483}
\bibfield{author}{\bibinfo{person}{Ang Li}, \bibinfo{person}{Shuaiwen~Leon Song}, \bibinfo{person}{Jieyang Chen}, \bibinfo{person}{Xu Liu}, \bibinfo{person}{Nathan Tallent}, {and} \bibinfo{person}{Kevin Barker}.} \bibinfo{year}{2018}\natexlab{b}.
\newblock \showarticletitle{Tartan: Evaluating Modern GPU Interconnect via a Multi-GPU Benchmark Suite}. In \bibinfo{booktitle}{\emph{2018 IEEE International Symposium on Workload Characterization (IISWC)}}. \bibinfo{pages}{191--202}.
\newblock
\href{https://doi.org/10.1109/IISWC.2018.8573483}{doi:\nolinkurl{10.1109/IISWC.2018.8573483}}


\bibitem[Li et~al\mbox{.}(2018a)]%
        {li2018parti}
\bibfield{author}{\bibinfo{person}{Jiajia Li}, \bibinfo{person}{Yuchen Ma}, {and} \bibinfo{person}{Richard Vuduc}.} \bibinfo{year}{2018}\natexlab{a}.
\newblock \showarticletitle{ParTI!: A parallel tensor infrastructure for multicore CPUs and GPUs}.
\newblock \bibinfo{journal}{\emph{A parallel tensor infrastructure for multicore CPUs and GPUs}} (\bibinfo{year}{2018}).
\newblock


\bibitem[Li et~al\mbox{.}(2019)]%
        {hicoo_repo}
\bibfield{author}{\bibinfo{person}{Jiajia Li}, \bibinfo{person}{Bora U\c{c}ar}, \bibinfo{person}{\"{U}mit~V. \c{C}ataly\"{u}rek}, \bibinfo{person}{Jimeng Sun}, \bibinfo{person}{Kevin Barker}, {and} \bibinfo{person}{Richard Vuduc}.} \bibinfo{year}{2019}\natexlab{}.
\newblock \bibinfo{title}{Efficient and Effective Sparse Tensor Reordering}.
\newblock
\urldef\tempurl%
\url{https://github.com/hpcgarage/ParTI}
\showURL{%
\tempurl}


\bibitem[Lin et~al\mbox{.}(2024)]%
        {10740878}
\bibfield{author}{\bibinfo{person}{Wenqing Lin}, \bibinfo{person}{Hemeng Wang}, \bibinfo{person}{Haodong Deng}, {and} \bibinfo{person}{Qingxiao Sun}.} \bibinfo{year}{2024}\natexlab{}.
\newblock \showarticletitle{ScalFrag: Efficient Tiled-MTTKRP with Adaptive Launching on GPUs}. In \bibinfo{booktitle}{\emph{2024 IEEE International Conference on Cluster Computing (CLUSTER)}}. \bibinfo{pages}{335--345}.
\newblock
\href{https://doi.org/10.1109/CLUSTER59578.2024.00036}{doi:\nolinkurl{10.1109/CLUSTER59578.2024.00036}}


\bibitem[McAuley(2021)]%
        {recommend_repo}
\bibfield{author}{\bibinfo{person}{Julian McAuley}.} \bibinfo{year}{2021}\natexlab{}.
\newblock \bibinfo{title}{Recommender Systems and Personalization Datasets}.
\newblock
\urldef\tempurl%
\url{https://cseweb.ucsd.edu/~jmcauley/datasets.html#}
\showURL{%
\tempurl}


\bibitem[Nguyen et~al\mbox{.}(2022a)]%
        {10.1145/3524059.3532363}
\bibfield{author}{\bibinfo{person}{Andy Nguyen}, \bibinfo{person}{Ahmed~E. Helal}, \bibinfo{person}{Fabio Checconi}, \bibinfo{person}{Jan Laukemann}, \bibinfo{person}{Jesmin~Jahan Tithi}, \bibinfo{person}{Yongseok Soh}, \bibinfo{person}{Teresa Ranadive}, \bibinfo{person}{Fabrizio Petrini}, {and} \bibinfo{person}{Jee~W. Choi}.} \bibinfo{year}{2022}\natexlab{a}.
\newblock \showarticletitle{Efficient, out-of-Memory Sparse MTTKRP on Massively Parallel Architectures}. In \bibinfo{booktitle}{\emph{Proceedings of the 36th ACM International Conference on Supercomputing}} (Virtual Event) \emph{(\bibinfo{series}{ICS '22})}. \bibinfo{publisher}{Association for Computing Machinery}, \bibinfo{address}{New York, NY, USA}, Article \bibinfo{articleno}{26}, \bibinfo{numpages}{13}~pages.
\newblock
\showISBNx{9781450392815}
\href{https://doi.org/10.1145/3524059.3532363}{doi:\nolinkurl{10.1145/3524059.3532363}}


\bibitem[Nguyen et~al\mbox{.}(2022b)]%
        {blco_github}
\bibfield{author}{\bibinfo{person}{Andy Nguyen}, \bibinfo{person}{Ahmed~E Helal}, \bibinfo{person}{Fabio Checconi}, \bibinfo{person}{Jan Laukemann}, \bibinfo{person}{Jesmin~Jahan Tithi}, \bibinfo{person}{Yongseok Soh}, \bibinfo{person}{Teresa Ranadive}, \bibinfo{person}{Fabrizio Petrini}, {and} \bibinfo{person}{Jee~W Choi}.} \bibinfo{year}{2022}\natexlab{b}.
\newblock \bibinfo{title}{Efficient, out-of-memory sparse MTTKRP on massively parallel architectures}.
\newblock
\urldef\tempurl%
\url{https://github.com/jeewhanchoi/blocked-linearized-coordinate}
\showURL{%
\tempurl}


\bibitem[Nisa et~al\mbox{.}(2019a)]%
        {10.1145/3295500.3356216}
\bibfield{author}{\bibinfo{person}{Israt Nisa}, \bibinfo{person}{Jiajia Li}, \bibinfo{person}{Aravind Sukumaran-Rajam}, \bibinfo{person}{Prasant~Singh Rawat}, \bibinfo{person}{Sriram Krishnamoorthy}, {and} \bibinfo{person}{P. Sadayappan}.} \bibinfo{year}{2019}\natexlab{a}.
\newblock \showarticletitle{An Efficient Mixed-Mode Representation of Sparse Tensors}. In \bibinfo{booktitle}{\emph{Proceedings of the International Conference for High Performance Computing, Networking, Storage and Analysis}} (Denver, Colorado) \emph{(\bibinfo{series}{SC '19})}. \bibinfo{publisher}{Association for Computing Machinery}, \bibinfo{address}{New York, NY, USA}, Article \bibinfo{articleno}{49}, \bibinfo{numpages}{25}~pages.
\newblock
\showISBNx{9781450362290}
\href{https://doi.org/10.1145/3295500.3356216}{doi:\nolinkurl{10.1145/3295500.3356216}}


\bibitem[Nisa et~al\mbox{.}(2019b)]%
        {mm_csf}
\bibfield{author}{\bibinfo{person}{Israt Nisa}, \bibinfo{person}{Jiajia Li}, \bibinfo{person}{Aravind Sukumaran-Rajam}, \bibinfo{person}{Prasant~Singh Rawat}, \bibinfo{person}{Sriram Krishnamoorthy}, {and} \bibinfo{person}{Ponnuswamy Sadayappan}.} \bibinfo{year}{2019}\natexlab{b}.
\newblock \bibinfo{title}{An Efficient Mixed-Mode Representation of Sparse Tensors}.
\newblock
\urldef\tempurl%
\url{https://github.com/isratnisa/MM-CSF}
\showURL{%
\tempurl}


\bibitem[Nisa et~al\mbox{.}(2019c)]%
        {8821030}
\bibfield{author}{\bibinfo{person}{Israt Nisa}, \bibinfo{person}{Jiajia Li}, \bibinfo{person}{Aravind Sukumaran-Rajam}, \bibinfo{person}{Richard Vuduc}, {and} \bibinfo{person}{P. Sadayappan}.} \bibinfo{year}{2019}\natexlab{c}.
\newblock \showarticletitle{Load-Balanced Sparse MTTKRP on GPUs}. In \bibinfo{booktitle}{\emph{2019 IEEE International Parallel and Distributed Processing Symposium (IPDPS)}}. \bibinfo{pages}{123--133}.
\newblock
\href{https://doi.org/10.1109/IPDPS.2019.00023}{doi:\nolinkurl{10.1109/IPDPS.2019.00023}}


\bibitem[{NVIDIA Corporation}(2024a)]%
        {nvidia_gpudirect}
\bibfield{author}{\bibinfo{person}{{NVIDIA Corporation}}.} \bibinfo{year}{2024}\natexlab{a}.
\newblock \bibinfo{title}{{GPUDirect}}.
\newblock
\urldef\tempurl%
\url{https://developer.nvidia.com/gpudirect}
\showURL{%
\tempurl}
\newblock
\shownote{Accessed: 2025-04-30}.


\bibitem[{NVIDIA Corporation}(2024b)]%
        {nvidia_NVLINK}
\bibfield{author}{\bibinfo{person}{{NVIDIA Corporation}}.} \bibinfo{year}{2024}\natexlab{b}.
\newblock \bibinfo{title}{{NVIDIA NVLink}}.
\newblock
\urldef\tempurl%
\url{https://www.nvidia.com/en-us/design-visualization/nvlink-bridges/}
\showURL{%
\tempurl}
\newblock
\shownote{Accessed: 2025-04-30}.


\bibitem[Owens et~al\mbox{.}(2008)]%
        {4490127}
\bibfield{author}{\bibinfo{person}{John~D. Owens}, \bibinfo{person}{Mike Houston}, \bibinfo{person}{David Luebke}, \bibinfo{person}{Simon Green}, \bibinfo{person}{John~E. Stone}, {and} \bibinfo{person}{James~C. Phillips}.} \bibinfo{year}{2008}\natexlab{}.
\newblock \showarticletitle{GPU Computing}.
\newblock \bibinfo{journal}{\emph{Proc. IEEE}} \bibinfo{volume}{96}, \bibinfo{number}{5} (\bibinfo{year}{2008}), \bibinfo{pages}{879--899}.
\newblock
\href{https://doi.org/10.1109/JPROC.2008.917757}{doi:\nolinkurl{10.1109/JPROC.2008.917757}}


\bibitem[Rappaz et~al\mbox{.}(2021)]%
        {10.1145/3460231.3474267}
\bibfield{author}{\bibinfo{person}{J\'{e}r\'{e}mie Rappaz}, \bibinfo{person}{Julian McAuley}, {and} \bibinfo{person}{Karl Aberer}.} \bibinfo{year}{2021}\natexlab{}.
\newblock \showarticletitle{Recommendation on Live-Streaming Platforms: Dynamic Availability and Repeat Consumption}. In \bibinfo{booktitle}{\emph{Proceedings of the 15th ACM Conference on Recommender Systems}} (Amsterdam, Netherlands) \emph{(\bibinfo{series}{RecSys '21})}. \bibinfo{publisher}{Association for Computing Machinery}, \bibinfo{address}{New York, NY, USA}, \bibinfo{pages}{390–399}.
\newblock
\showISBNx{9781450384582}
\href{https://doi.org/10.1145/3460231.3474267}{doi:\nolinkurl{10.1145/3460231.3474267}}


\bibitem[Smith et~al\mbox{.}(2017)]%
        {frosttdataset}
\bibfield{author}{\bibinfo{person}{Shaden Smith}, \bibinfo{person}{Jee~W. Choi}, \bibinfo{person}{Jiajia Li}, \bibinfo{person}{Richard Vuduc}, \bibinfo{person}{Jongsoo Park}, \bibinfo{person}{Xing Liu}, {and} \bibinfo{person}{George Karypis}.} \bibinfo{year}{2017}\natexlab{}.
\newblock \bibinfo{booktitle}{\emph{{FROSTT}: The Formidable Repository of Open Sparse Tensors and Tools}}.
\newblock
\urldef\tempurl%
\url{http://frostt.io/}
\showURL{%
\tempurl}


\bibitem[Srivastava et~al\mbox{.}(2020)]%
        {9065579}
\bibfield{author}{\bibinfo{person}{Nitish Srivastava}, \bibinfo{person}{Hanchen Jin}, \bibinfo{person}{Shaden Smith}, \bibinfo{person}{Hongbo Rong}, \bibinfo{person}{David Albonesi}, {and} \bibinfo{person}{Zhiru Zhang}.} \bibinfo{year}{2020}\natexlab{}.
\newblock \showarticletitle{Tensaurus: A Versatile Accelerator for Mixed Sparse-Dense Tensor Computations}. In \bibinfo{booktitle}{\emph{2020 IEEE International Symposium on High Performance Computer Architecture (HPCA)}}. \bibinfo{pages}{689--702}.
\newblock
\href{https://doi.org/10.1109/HPCA47549.2020.00062}{doi:\nolinkurl{10.1109/HPCA47549.2020.00062}}


\bibitem[Wijeratne et~al\mbox{.}(2021)]%
        {9622851}
\bibfield{author}{\bibinfo{person}{Sasindu Wijeratne}, \bibinfo{person}{Rajgopal Kannan}, {and} \bibinfo{person}{Viktor Prasanna}.} \bibinfo{year}{2021}\natexlab{}.
\newblock \showarticletitle{Reconfigurable Low-latency Memory System for Sparse Matricized Tensor Times Khatri-Rao Product on FPGA}. In \bibinfo{booktitle}{\emph{2021 IEEE High Performance Extreme Computing Conference (HPEC)}}. \bibinfo{pages}{1--7}.
\newblock
\href{https://doi.org/10.1109/HPEC49654.2021.9622851}{doi:\nolinkurl{10.1109/HPEC49654.2021.9622851}}


\bibitem[Wijeratne et~al\mbox{.}(2023a)]%
        {wijeratne2023dynasor}
\bibfield{author}{\bibinfo{person}{Sasindu Wijeratne}, \bibinfo{person}{Rajgopal Kannan}, {and} \bibinfo{person}{Viktor Prasanna}.} \bibinfo{year}{2023}\natexlab{a}.
\newblock \showarticletitle{Dynasor: A Dynamic Memory Layout for Accelerating Sparse MTTKRP for Tensor Decomposition on Multi-core CPU}. In \bibinfo{booktitle}{\emph{2023 IEEE 35th International Symposium on Computer Architecture and High Performance Computing (SBAC-PAD)}}. IEEE, \bibinfo{pages}{23--33}.
\newblock


\bibitem[Wijeratne et~al\mbox{.}(2024)]%
        {wijeratne2024sparse}
\bibfield{author}{\bibinfo{person}{Sasindu Wijeratne}, \bibinfo{person}{Rajgopal Kannan}, {and} \bibinfo{person}{Viktor Prasanna}.} \bibinfo{year}{2024}\natexlab{}.
\newblock \showarticletitle{Sparse MTTKRP Acceleration for Tensor Decomposition on GPU}. In \bibinfo{booktitle}{\emph{Proceedings of the 21st ACM International Conference on Computing Frontiers}}. \bibinfo{pages}{88--96}.
\newblock


\bibitem[Wijeratne et~al\mbox{.}(2023b)]%
        {10.1145/3543622.3573179}
\bibfield{author}{\bibinfo{person}{Sasindu Wijeratne}, \bibinfo{person}{Ta-Yang Wang}, \bibinfo{person}{Rajgopal Kannan}, {and} \bibinfo{person}{Viktor Prasanna}.} \bibinfo{year}{2023}\natexlab{b}.
\newblock \showarticletitle{Accelerating Sparse MTTKRP for Tensor Decomposition on FPGA}. In \bibinfo{booktitle}{\emph{Proceedings of the 2023 ACM/SIGDA International Symposium on Field Programmable Gate Arrays}} (Monterey, CA, USA) \emph{(\bibinfo{series}{FPGA '23})}. \bibinfo{publisher}{Association for Computing Machinery}, \bibinfo{address}{New York, NY, USA}, \bibinfo{pages}{259–269}.
\newblock
\showISBNx{9781450394178}
\href{https://doi.org/10.1145/3543622.3573179}{doi:\nolinkurl{10.1145/3543622.3573179}}


\bibitem[Zeller(2011)]%
        {ansorge2022programming}
\bibfield{author}{\bibinfo{person}{Cyril Zeller}.} \bibinfo{year}{2011}\natexlab{}.
\newblock \showarticletitle{CUDA C/C++ Basics}.
\newblock  (\bibinfo{year}{2011}).
\newblock


\end{thebibliography}


\end{document}